\documentclass[12pt]{article}
\usepackage{epsf}

\begin{document}

\newcommand{\gtrsim}{\mathop{}_{\textstyle \sim}^{\textstyle >}}
\newcommand{\lesssim}{\mathop{}_{\textstyle \sim}^{\textstyle <} }

\newcommand{\rem}[1]{{\bf #1}}

\renewcommand{\theequation}{\thesection.\arabic{equation}}

\renewcommand{\thefootnote}{\fnsymbol{footnote}}
\setcounter{footnote}{0}

\begin{titlepage}

\def\thefootnote{\fnsymbol{footnote}}

\begin{center}

\hfill TU-799\\
\hfill September, 2007\\

\vskip .75in

{\Large \bf 

Decay of Scalar Condensation \\
in Quantum Field Theory

}

\vskip .75in

{\large
Shigeki Matsumoto$^{\rm (a,b)}$ and Takeo Moroi$^{\rm (a)}$
}

\vskip 0.25in

$^{\rm (a)}${\em Department of Physics, Tohoku University,
Sendai 980-8578, Japan}

\vskip 0.25in

$^{\rm (b)}${\em
Tohoku University
International Advanced Research and Education Organization,\\
Institute for International Advanced Interdisciplinary Research,\\
Sendai 980-8578, Japan}

\end{center}
\vskip .5in

\begin{abstract}

  We consider decay processes of scalar-field condensation in the
  framework of well-established quantum field theory.  We postulate
  that the quantum state corresponding to the scalar-field
  condensation is so-called coherent state with discussing the
  validity of such a treatment.  We show that, by using the unitarity
  relation of the scattering matrix, decay rate of the coherent state
  is systematically calculated.  We apply our procedure to derive
  explicit formulae of decay rates for two cases: (i) we study the
  case where the scalar condensation decays into a pair of scalar
  particles and show that our formalism reproduces the results
  obtained from the parametric-resonance analysis, and (ii) we also
  calculate the decay rate when the coherent state decays via anomaly.

\end{abstract}

\end{titlepage}

\renewcommand{\thepage}{\arabic{page}}
\setcounter{page}{1}
\renewcommand{\thefootnote}{\#\arabic{footnote}}
\setcounter{footnote}{0}

\section{Introduction}
\label{sec:introduction}
\setcounter{equation}{0}

Scalar-field condensations play very crucial roles in various places
in cosmology.  Probably the most important example is the inflaton
field which is necessary for inflation \cite{Guth:1980zm,Sato:1980yn}.
In particular, in the slow-roll inflation models
\cite{Linde:1981mu,Albrecht:1982wi}, energy density of the inflaton
condensation provides the energy density to realize (quasi-) de-Sitter
universe during inflation.  After inflation, the inflaton oscillates
around the minimum of the potential and decays into standard-model
particles to reheat the universe realizing hot big-bang cosmology.
This class of inflation model not only solves flatness and horizon
problems in cosmology but also provides density fluctuation consistent
with the Wilkinson Microwave Anisotropy Probe data
\cite{Spergel:2006hy}.  In addition, it has been pointed out that the
density fluctuation may arise from late-decaying scalar field other
than the inflaton, which is called ``curvaton''
\cite{Enqvist:2001zp,Lyth:2001nq,Moroi:2001ct}.  Other important
example is the Affleck-Dine field for baryogenesis
\cite{Affleck:1984fy}.  In low-energy supersymmetric models, there
exist scalar fields, i.e., scalar partners of quarks, which have
baryon number.  If some of those fields acquire non-vanishing
amplitudes in the early universe, non-vanishing baryon number may be
imprinted into the motion of the scalar-quark condensations due to
baryon-number violating operators at an ultra-high energy scale. Such
a scenario is one of the most attractive scenario to generate large
enough baryon asymmetry of the universe.

All of these exotic scalar-field condensations (i.e, inflaton,
curvaton, Affleck-Dine field, and so on) oscillate around the minimum
of the potential at some stage of the evolution of the universe, and
eventually decay into standard-model particles for the cosmological
history consistent with observations.  Thus, it is important to
understand how the scalar-field condensation decays from the view
point of the quantum field theory.

The main concern of this paper is to discuss how the oscillating
scalar-field condensation decays into other states.  Around the
minimum, the potential of the scalar field $\varphi$ is well estimated
by a parabolic one
\begin{eqnarray}
 V = \frac{1}{2} m_\varphi^2 \varphi^2.
\end{eqnarray}
Neglecting the effect of the cosmic expansion for simplicity, the
solution to the classical equation of motion is given by
\begin{eqnarray}
 \varphi = A_\varphi \cos m_\varphi t.
  \label{cos(mt)}
\end{eqnarray}
In this case, the energy density of the condensation is given by
$\rho_\varphi=\frac{1}{2}m_\varphi^2A_\varphi^2$.  One should
understand how the energy density stored in the oscillation of
$\varphi$ is converted to the that of radiation.  In the simplest
approach, the decay rate of the scalar-field condensation is estimated
from the decay rate of single scalar field (in the vacuum): the energy
density of the scalar field in the condensation is approximated to
decrease as
\begin{eqnarray}
 \frac{d \rho_\varphi}{d t} = 
  - \gamma_\varphi \rho_\varphi,
  \label{dot(rho)}
\end{eqnarray}
where $\gamma_\varphi$ is the decay rate of $\varphi$ in the vacuum.
However, it has been also pointed out that, when the scalar field
$\varphi$ is oscillating like Eq.\ (\ref{cos(mt)}), wave functions of
fields which couple to $\varphi$ are modified.  Consequently, the
``decay rate'' of the scalar field in the condensation may be
significantly different from the one obtained from the field theory in
the vacuum.  In particular, in some of the cases, instability band may
arise in the wave function of the final-state particles, which results
in catastrophic particle production (so-called parametric resonance)
\cite{Kofman:1994rk,Shtanov:1994ce,Yoshimura:1995gc,Kofman:1997yn}.
Since the decay of scalar-field condensation is very important, it is
desirable to have a deep understanding of the decay processes of
scalar-field oscillations in the framework of well-established quantum
field theory.

In this paper, we consider decay processes of scalar condensations in
the framework of the quantum field theory.  In our analysis, we
neglect the effects of cosmic expansion as a first step to understand
the behavior of the scalar condensation.  We postulate that the scalar
condensation corresponds to the so-called coherent state
$|\varphi\rangle$ in the quantum field theory; justification of such a
treatment will be also discussed.  Then, we will show that the decay
rate of the coherent state can be systematically calculated by using
the unitarity relation of the scattering matrix ($S$-matrix).  By
using the fact that the decay rate of the coherent state is
proportional to the imaginary part of the so-called $T$-matrix element
$\Im[\langle\varphi |\hat{\cal T}|\varphi\rangle]$, we formulate the
calculation of the decay rate of the coherent state.  We apply our
procedure to the case where the scalar-field condensation is coupled
to a real scalar field and calculate the decay rate for such a case.
We will see that our procedure gives the same decay rate as that
obtained by the discussion of parametric-resonance.  We discuss when
Eq.\ (\ref{dot(rho)}) is justified and how the instability band for
parametric resonance arises in our framework.  We also calculate the
decay rate of scalar condensation which decays into gauge-boson pair
via anomaly.

Organization of this paper is as follows.  In Section\
\ref{sec:basic}, we first derive basic formulae which are used in the
calculation of the decay rate of the coherent state.  In particular,
we define the coherent state $|\varphi\rangle$ in the quantum field
theory and present important properties of $|\varphi\rangle$.  Then,
we explain how the decay rate of the state $|\varphi\rangle$ is
obtained.  In Section\ \ref{sec:a-term}, we consider the case where
the condensation couples to a real scalar field $\chi$ via three-point
interaction.  Decay of the coherent state via the interaction induced
by the anomaly is discussed in Section\ \ref{sec:anomaly}.  In
Section\ \ref{sec:summary}, we summarize our results.

\section{Basic Formulae}
\label{sec:basic}
\setcounter{equation}{0}

In this section, we introduce basic formulae used in our analysis.  We
discuss how the condensation of scalar field decays via some
interaction.  We assume that the interaction of the scalar field is
weak enough.

Total decay rate of any state can be related to the imaginary part of
the scattering-matrix element due to the $S$-matrix unitarity.  
Let us denote the $S$-matrix as
\begin{eqnarray}
  \hat{\cal S} = \hat{\bf 1} + i \hat{\cal T},
\end{eqnarray}
where $\hat{\bf 1}$ and $\hat{\cal T}$ are the unit operator and the
so-called $T$-matrix, respectively.  (Here and hereafter, the ``hat''
is used for operators.)  Then, from the unitarity of the $S$-matrix,
the following relation holds:
\begin{eqnarray}
  \hat{\cal T}^\dagger \hat{\cal T} =
  i (\hat{\cal T}^\dagger - \hat{\cal T}).
\end{eqnarray}
This relation is important for our analysis.

We expect that there exists a quantum state $|\varphi\rangle$ which
describes the state with scalar condensation.  (Details about
$|\varphi\rangle$ will be explained below.)  The probability of the
state $|\varphi\rangle$ decaying into all the possible final states is
related to the imaginary part of the $T$-matrix element as
\begin{eqnarray}
  {\rm Prob} ( | \varphi \rangle \rightarrow {\rm all} ) =
  \sum_{f} \left| \langle f | \hat{\cal T} | \varphi \rangle \right|^2
  = 2 \Im
  \left[ \langle \varphi | \hat{\cal T} | \varphi \rangle \right].
\end{eqnarray}

Now, we consider what the state $|\varphi\rangle$ is.  We first
quantize the field operator using the free part of the Lagrangian,
then treat the interaction terms as perturbations.  We denote the free
part of the Lagrangian of the real scalar field $\varphi$ as
\begin{eqnarray}
  {\cal L} = \frac{1}{2} \partial_\mu \varphi \partial^\mu \varphi
  - \frac{1}{2} m_\varphi^2 \varphi^2.
\end{eqnarray}
In our analysis, we use the box normalization of the wave functions
with the volume $L^3$.  Then, the field operator is given by
\begin{eqnarray}
  \hat{\varphi} (x) = \sum_{\bf k}
  \frac{1}{\sqrt{2E_{\bf k} L^3}}
  \left( \hat{a}_{\bf k} e^{-ikx} + \hat{a}_{\bf k}^\dagger e^{ikx} \right),
  \label{varphi(real)}
\end{eqnarray}
where $E_{\bf k}\equiv\sqrt{{\bf k}^2+m_\varphi^2}$.  The annihilation
and creation operators satisfy the following commutation relations:
\begin{eqnarray}
  [\hat{a}_{\bf k}, \hat{a}_{\bf k'}^\dagger] = \delta_{{\bf k}, {\bf k'}},
\end{eqnarray}
while $\hat{a}_{\bf k}$ and $\hat{a}_{\bf k'}$ (as well as
$\hat{a}_{\bf k}^\dagger$ and $\hat{a}_{\bf k'}^\dagger$) commute.

We postulate that the quantum state describing the scalar-field
condensation is the coherent state, which is given by
\begin{eqnarray}
  | \varphi \rangle \equiv
  e^{-|C_\varphi|^2/2}
  e^{C_\varphi \hat{a}_{\bf 0}^\dagger} | 0 \rangle,
\end{eqnarray}
where $| 0 \rangle$ is the vacuum state satisfying $\hat{a}_{\bf k}| 0
\rangle=0$.  Notice that the state $|\varphi\rangle$ is properly
normalized: $\langle\varphi |\varphi\rangle=1$.  In addition,
importantly, the state $|\varphi\rangle$ is an eigenstate of the
annihilation operator $\hat{a}_{\bf 0}$:
\begin{eqnarray}
  \hat{a}_{\bf 0} | \varphi \rangle = 
  C_\varphi | \varphi \rangle.
  \label{a|phi>}
\end{eqnarray}
We can also see that
\begin{eqnarray}
  \varphi (x) \equiv
  \langle \varphi | \hat{\varphi} (x) | \varphi \rangle
  = 
  \varphi_- (x) + \varphi_+ (x),
  \label{<varphi>}
\end{eqnarray}
where
\begin{eqnarray}
  \varphi_- (x) \equiv
  \frac{1}{2}
  A_\varphi e^{-im_\varphi t}, ~~~
  \varphi_+ (x) \equiv
  \frac{1}{2} A_\varphi^* e^{im_\varphi t},
\end{eqnarray}
with
\begin{eqnarray}
  A_\varphi = C_\varphi \sqrt{\frac{2}{m_\varphi L^3}}.
\end{eqnarray}
One can easily see that, for the coherent state $|\varphi\rangle$, the
expectation value of the field operator follows the trajectory of the
solution to the classical wave equation.  Thus, we expect that the
coherent state $|\varphi\rangle$ corresponds to the quantum state
where the scalar field is under oscillation.

In calculating $T$-matrix elements, it is necessary to calculate the
expectation values of time-ordered products of field operators.  By
using the Wick's theorem, such products are calculated as
\begin{eqnarray}
  \langle \varphi | T \prod_i \hat{\varphi} (x_i) | \varphi \rangle
  &=& 
  \langle \varphi | N \prod_i \hat{\varphi} (x_i) | \varphi \rangle
  + (\mbox{all the possible contractions})
  \nonumber \\
  &=& 
  \prod_i \varphi (x_i)
  + (\mbox{all the possible contractions}),
\end{eqnarray}
where the symbol $T$ here denotes the time-ordering while $N$ is for
normal-ordering.  In addition, in the second equality, we have used
Eq.\ (\ref{a|phi>}).  Even in more complicated cases, we obtain
\begin{eqnarray}
  \langle \varphi | T \prod_i f_i 
   \left( \hat{\varphi} (x_i) \right) 
   | \varphi \rangle
  &=& 
  \Big\langle \varphi \Big| T \prod_i 
   \sum_n \frac{1}{n!}
   \left[ \frac{d^n f_i}{d \varphi^n} \right]_{\varphi (x_i)}
   \left(  \hat{\varphi} (x_i) - \varphi (x_i) \right)^n
  \Big| \varphi \Big\rangle
  \nonumber \\
  &=& 
  \prod_i f_i \left( \varphi (x_i) \right) 
  + (\mbox{all the possible contractions}).
  \label{Wick's2}
\end{eqnarray}
Here, we expand $f_i (\hat{\varphi} (x_i))$ around $\hat{\varphi}
(x_i)=\varphi (x_i)$.  Then,
$\langle\varphi|(\hat{\varphi}(x)-\varphi(x))|\varphi\rangle=0$, and
Eq.\ (\ref{Wick's2}) is applicable even when the function
$f_i(\varphi)$ is singular at $\varphi=0$.

In the following, we will not consider the processes in which
$\varphi$ is produced due to the decay of the coherent state.  (The
inclusion of such processes is straightforward.)  In such a case,
propagator of $\varphi$ does not show up in the calculation and the
field operator $\hat{\varphi}(x)$ can be simply replaced by the
expectation value $\varphi(x)$, in which the contraction terms are
irrelevant.

It is also instructive to calculate the expectation values of
energy-density operator as
\begin{eqnarray}
 \rho_\varphi
  = 
  L^{-3} \langle \varphi | 
  \sum_{\bf k} E_{\bf k} 
  \hat{a}_{\bf k}^\dagger \hat{a}_{\bf k}
  | \varphi \rangle
  = 
  \frac{1}{2} m_\varphi^2 |A_\varphi|^2,
\end{eqnarray}
while the expectation value of the number density is also given by
\begin{eqnarray}
  n_\varphi = 
  L^{-3} \langle \varphi | 
  \sum_{\bf k}
  \hat{a}_{\bf k}^\dagger \hat{a}_{\bf k}
  | \varphi \rangle
  = \frac{1}{2} m_\varphi |A_\varphi|^2.
\end{eqnarray}

For the complex scalar field (which we denote as $\phi$), similar
argument applies.  We define the field operator for the complex scalar
field as
\begin{eqnarray}
  \hat{\phi} (x) = \sum_{\bf k}
  \frac{1}{\sqrt{2E_{\bf k} L^3}}
  \left( \hat{a}_{\bf k} e^{-ikx} + \hat{b}_{\bf k}^\dagger e^{ikx} \right),
  \label{varphi(complex)}
\end{eqnarray}
where $\hat{a}_{\bf k}$ and $\hat{b}_{\bf k}$ ($\hat{a}_{\bf p}^\dagger$
and $\hat{b}_{\bf k}^\dagger$) are annihilation (creation) operators.
The coherent state for the complex field is given by
\begin{eqnarray}
  | \phi \rangle \equiv
  e^{-(|C_\phi|^2 + |C_{\bar{\phi}}|^2)/2}
  e^{C_\phi \hat{a}_{\bf 0}^\dagger 
    + C_{\bar{\phi}} \hat{b}_{\bf 0}^\dagger} | 0 \rangle,
  \label{|complex>}
\end{eqnarray}
and
\begin{eqnarray}
  \phi (x)
  \equiv \langle \phi | \hat{\phi} (x) | \phi \rangle
  = A_\phi e^{-im_\phi t} + A_{\bar{\phi}}^* e^{im_\phi t},
  \label{<complex-phi>}
\end{eqnarray}
where
\begin{eqnarray}
  A_\phi = \frac{C_\phi}{\sqrt{2m_\phi L^3}},~~~
  A_{\bar{\phi}} = \frac{C_{\bar{\phi}}}{\sqrt{2m_\phi L^3}}.
\end{eqnarray}
Energy density of this state is given by
\begin{eqnarray}
  \rho_\phi = 2 m_\phi^2 ( |A_\phi|^2 + |A_{\bar{\phi}}|^2 ),
\end{eqnarray}
while we can also calculate the expectation values of the number
densities of particle $\phi$ and anti-particle $\bar{\phi}$ as
\begin{eqnarray}
  n_\phi &\equiv&
  L^{-3} 
  \langle \phi | \sum_{\bf k} a_{\bf k}^\dagger a_{\bf k} | \phi \rangle
  = 2 m_\phi |A_\phi|^2,
  \label{n_phi}\\
  n_{\bar{\phi}} &\equiv&
  L^{-3} 
  \langle \phi | \sum_{\bf k} b_{\bf k}^\dagger b_{\bf k} | \phi \rangle
  = 2 m_\phi |A_{\bar{\phi}}|^2.
  \label{n_phibar}
\end{eqnarray}
We can see that the number densities of $\phi$ and its anti-particle
are proportional to $|A_\phi|^2$ and $|A_{\bar{\phi}}|^2$,
respectively.  Thus, when $|A_\phi|>|A_{\bar{\phi}}|$
($|A_\phi|<|A_{\bar{\phi}}|$), $\phi$ is more (less) abundant than
$\bar{\phi}$ in the condensation.  It should be also noted that the
function $\phi(x)$ given in Eq.\ (\ref{<complex-phi>}) gives an
elliptical trajectory on the complex $\phi$-plane.  When $A_\phi=0$ or
$A_{\bar{\phi}}=0$, the trajectory becomes a circle and, when
$|A_\phi|=|A_{\bar{\phi}}|$, the trajectory becomes a straight line.

\section{Decay into Scalar Fields}
\label{sec:a-term}
\setcounter{equation}{0}

\subsection{Setup}

First, we consider the simplest case where the scalar field $\varphi$
couples only to the real scalar field $\chi$ via the interaction
\begin{eqnarray}
  {\cal L}_{\rm int} = -\frac{1}{2} \mu \varphi \chi^2,
  \label{A-term}
\end{eqnarray}
with $\mu$ being the coupling constant.  With this interaction, the
decay rate of single particle in the vacuum is given by
\begin{eqnarray}
  \gamma_{\varphi\rightarrow\chi\chi} = \frac{\mu^2}{32\pi m_\varphi} 
  \sqrt{1 - \left( \frac{4m_\chi^2}{m_\varphi^2} \right) }.
  \label{Gamma(1P)}
\end{eqnarray}
In this section, with the interaction given in Eq.\ (\ref{A-term}), we
discuss how the coherent state decays.  

As discussed in the previous section, the decay rate of the coherent
state can be related to the imaginary part of the diagonal element of
the $T$-matrix $\langle\varphi |\hat{\cal T}|\varphi\rangle$.
Importantly, $\langle\varphi |\hat{\cal T}|\varphi\rangle$ is obtained
by calculating loop diagrams in the quantum field theory.

At the one-loop level, in other words, neglecting the fluctuation of
$\varphi$, $\langle\varphi |\hat{\cal T}| \varphi\rangle$ is expressed
as
\begin{eqnarray}
  \langle\varphi |\hat{\cal T}| \varphi\rangle
  \equiv 
  \sum_{p=1}^{\infty} \sum_{{\cal F}^{(2p)}}
  {\cal T}^{{\cal F}^{(2p)}},
\end{eqnarray}
where the summation over ${\cal F}^{(2p)}$ is for all the possible
Feynman diagrams with $2p$ external $\varphi$.  With $p$ being fixed,
one can find
\begin{eqnarray}
 \sum_{{\cal F}^{(2p)}}
  i {\cal T}^{{\cal F}^{(2p)}}
  =
  \frac{1}{(2p)!} \left( \frac{i \mu}{2} \right)^{2p}
  \Big\langle 0 \Big| T
  \left[ \int d^4 x \left\{
    \varphi_- (x) + \varphi_+ (x)
    \right\}
    \hat{\chi} (x) \hat{\chi} (x) \right]^{2p}
  \Big| 0 \Big\rangle,
  \label{T^F}
\end{eqnarray}
where $\varphi_-$ and $\varphi_+$ show up when the field operator
$\hat{\varphi}$ is contracted with the creation operator in
$|\varphi\rangle$ and the annihilation operator in $\langle\varphi|$,
respectively.  In order to obtain non-vanishing results, we need to pick
up same number of $\varphi_-$ and $\varphi_+$.  

\begin{figure}[t]
 \centerline{\epsfxsize=0.4\textwidth\epsfbox{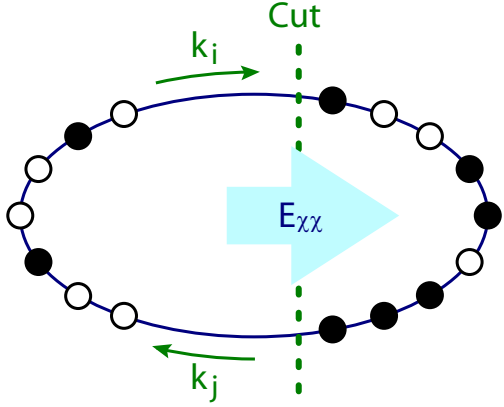}}
 \caption{\small An example of the Feynman diagram which gives rise to
   the imaginary part of $\langle\varphi|\hat{\cal T}|\varphi\rangle$.
   The white dot $\circ$ indicates $\varphi_-$ insertion while the
   black dot $\bullet$ is for $\varphi_+$ insertion.  In this example,
   $p=8$ and, with the cut shown in the figure, $N_\varphi=3$.
   (Notice that other cuts are also possible with this diagram.)}
\label{fig:loop_general}
\end{figure}

The right-hand side of Eq.\ (\ref{T^F}) contains contribution from
various Feynman diagrams because $\varphi_-$ and $\varphi_+$ can be
ordered in many ways.  In Fig.\ \ref{fig:loop_general}, we show a
typical diagram (after imposing the cut).  In our notation, we
represent the insertion of $\varphi_-$ by the white dot $\circ$ while
$\varphi_+$ by the black dot $\bullet$; non-vanishing diagrams have
$p$ black and $p$ white dots.  Internal lines are $\chi$-propagator.

Expectation value of $\hat{\cal T}$ for the given diagram ${\cal
F}^{(2p)}$ (with $2p$ external $\varphi$ insertion) is given by
\begin{eqnarray}
  i {\cal T}^{{\cal F}^{(2p)}}
  =
  L^3 T
  S_{\cal F}
  \left| \frac{\mu A_\varphi}{2} \right|^{2p}
  \int \frac{d^4 \tilde{k}}{(2\pi)^4}
  \prod_{I=1}^{2p}
  ( k_I^2 - m^2_\chi + i 0^+ )^{-2},
  \label{T(2p)}
\end{eqnarray}
where $S_{\cal F}$ is the symmetry factor and
\begin{eqnarray}
  L^3 T = \int d^4 x.
\end{eqnarray}
In addition, the momentum flowing on the $I$-th propagator is given by
\begin{eqnarray}
 k_I \equiv \tilde{k} + \sum_{J=1}^{I} \varepsilon_{J} Q_\varphi,
\end{eqnarray}
with 
\begin{eqnarray}
  Q_\varphi \equiv (m_\varphi, {\bf 0}),
\end{eqnarray}
and $\varepsilon_J=1$ and $-1$ if the $J$-th $\varphi$ insertion is
$\varphi_-$ and $\varphi_+$, respectively.

The imaginary part of ${\cal T}^{{\cal F}^{(2p)}}$ is obtained by
cutting two propagators (see Fig.\ \ref{fig:loop_general}), which
corresponds to the replacements of those two propagators by the
$\delta$-functions (with a relevant numerical factor):
\begin{eqnarray}
 \Im\left[ {\cal T}^{{\cal F}^{(2p)}} \right]
  = 
  \lim_{\xi\rightarrow m_\chi^2}
  \sum_{i=1}^{2p-1} \sum_{j=i+1}^{2p}
  \Im \left[ {\cal T}^{{\cal F}^{(2p)}}_{i,j} (\xi) \right],
  \label{Im(T)}
\end{eqnarray}
where
\begin{eqnarray}
  \Im \left[ {\cal T}^{{\cal F}^{(2p)}}_{i,j} (\xi) \right]
  \equiv
  2\pi^2 L^3 T
  S_{\cal F}
  \left| \frac{\mu A_\varphi}{2} \right|^{2p}
  \int \frac{d^4 \tilde{k}}{(2\pi)^4}
  \delta ( k_i^2 - \xi_i ) 
  \delta ( k_j^2 - \xi_j ) 
  \prod_{I\neq i,j}
  ( k_I^2 - \xi_I )^{-1}.
\end{eqnarray}
(For details, see Appendix \ref{app:im}.)  $\Im[ {\cal T}^{{\cal
F}^{(2p)}}_{i,j}]$ is the contribution from the diagram in which cut is
on $i$- and $j$-th propagators.  For $\Im[ {\cal T}^{{\cal
F}^{(2p)}}_{i,j}]$, we define the energy flow from one side of the cut
to the other, which we denote $E_{\chi\chi}$; with the following
non-negative integer:
\begin{eqnarray}
  N_\varphi = 
  \left| \sum_{I=i+1}^{j} \varepsilon_I \right|,
\end{eqnarray}
the energy flow is given by
\begin{eqnarray}
  E_{\chi\chi} = N_\varphi m_\varphi.
\end{eqnarray}
Notice that $\Im [ {\cal T}^{{\cal F}^{(2p)}}_{i,j} ]$ contributes
only to the decay rate of the process in which $N_\varphi$ of
$\varphi$ in the condensation simultaneously annihilate into two
$\chi$ because we neglect the fluctuations of the $\varphi$ field.
At the perturbative level, such a decay process is kinematically
allowed when $E_{\chi\chi}>2m_\chi$.

\subsection{Small amplitude limit}

\begin{figure}[t]
 \centerline{\epsfxsize=0.4\textwidth\epsfbox{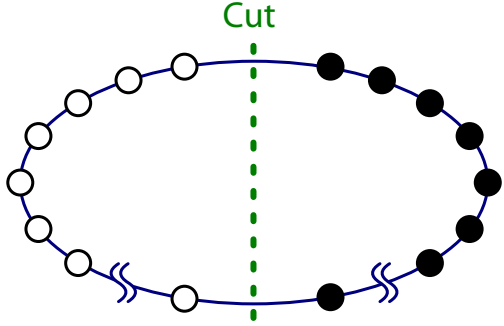}}
 \caption{\small Feynman diagram which gives the leading-order
   contribution to the decay rate in the small amplitude limit. The
   white dot $\circ$ indicates $\varphi_-$ insertion while the black
   dot $\bullet$ is for $\varphi_+$ insertion.}
 \label{fig:loop_leading}
\end{figure}

In this and the next subsections, we concentrate on the case where the
amplitude of $\varphi$ is small.  In this case, the leading-order
contribution to the $E_\varphi=N_\varphi m_\phi$ mode is from the
diagram with $2N_\varphi$ external $\varphi$ with $\varphi_-$ and
$\varphi_+$ being completely separated by the cut.  (See Fig.\
\ref{fig:loop_leading}.)  Concentrating on such a leading-order
diagram, the imaginary part of the $T$-matrix element is given by
\begin{eqnarray}
 \Im \left[ {\cal T}^{(N_\varphi)}_{\rm Leading} \right]
  &=& 
  L^3 T
  \pi^2 \left| \frac{\mu A_\varphi}{2} \right|^{2N_\varphi}
  \int \frac{d^4\tilde{k}}{(2\pi)^4}
  \delta \left( \tilde{k}^2 - m_\chi^2 \right)
  \delta 
  \left( ( \tilde{k} - N_\varphi Q_\varphi )^2 - m_\chi^2 \right)
  \nonumber \\ &&
  \prod_{I=1}^{N_\varphi-1} 
  \left[ ( \tilde{k} - I Q_\varphi )^2
  - m_\chi^2 \right]^{-2}.
\end{eqnarray}
Constraints from the $\delta$-functions give
$\tilde{k}Q_\varphi=\frac{1}{2}N_\varphi m_\varphi^2$.  Thus,
\begin{eqnarray}
 \prod_{I=1}^{N_\varphi-1} 
  \left[ ( \tilde{k} - I Q_\varphi )^2
   - m_\chi^2 \right]^{-1}
  \rightarrow
  m_\varphi^{ -4( N_\varphi - 1 ) } 
  \left[ ( N_\varphi - 1 )! \right]^{-4},
\end{eqnarray}
and hence
\begin{eqnarray}
 \Im \left[ {\cal T}^{(N_\varphi)}_{\rm Leading} \right]
  &=&
  L^3 T
  \frac{\beta_{N_\varphi}}{32\pi} \frac{m_\varphi^4}{[ (N_\varphi-1)! ]^4}
  \left| \frac{\mu A_\varphi}{2 m_\varphi^2}  \right|^{2N_\varphi},
\end{eqnarray}
where, for $N_\varphi m_\varphi >2m_\chi$, the velocity
$\beta_{N_\varphi}$ is given by
\begin{eqnarray}
  \beta_{N_\varphi} \equiv
  \sqrt{1 - \frac{4m_\chi^4}{N_\varphi^2 m_\varphi^2}},
  \label{beta_N}
\end{eqnarray}
while $\beta_{N_\varphi}=0$ 
for $N_\varphi m_\varphi \leq 2m_\chi$.

The decay rate of the coherent state $|\varphi\rangle$ per unit volume
is evaluated as
\begin{eqnarray}
 (\mbox{Decay rate per unit volume}) =
  \frac{{\rm Prob} (|\varphi\rangle \rightarrow {\rm all})}
  {L^3 T},
\end{eqnarray}
and hence is given by
\begin{eqnarray}
 \Gamma^{(N_\varphi)}_{\rm Leading} = 
  \frac{\beta_{N_\varphi}}{16\pi} \frac{m_\varphi^4}{[ (N_\varphi-1)! ]^4}
  \left| \frac{\mu A_\varphi}{2 m_\varphi^2}  \right|^{2N_\varphi}.
  \label{Gamma^N}
\end{eqnarray}
The above expression is consistent with the result given in the study
of the parametric resonance \cite{Yoshimura:1995gc}.  In addition, the
decay rate for the $N_\varphi=1$ mode is related to the decay rate of
single particle, which is given in Eq.\ (\ref{Gamma(1P)}), as
\begin{eqnarray}
 \Gamma^{(N_\varphi=1)}_{\rm Leading} = 
  n_\varphi \gamma_{\varphi\rightarrow\chi\chi}.
\end{eqnarray}
Thus, when the amplitude of $\varphi$ is small, decay of the coherent
state can be treated as the decay of individual particles in the
condensation, which justifies the conventional treatment of the decay
processes of scalar condensations.

We also comment here that the decay rate given in Eq.\ (\ref{Gamma^N})
is also derived from the tree-level calculation of the $\chi\chi$ pair
creation rate in the external oscillating $\varphi$ field;
$\Gamma^{(N_\varphi)}_{\rm Leading}$ is equal to the production rate
of $\chi\chi$ pair per unit volume with total energy of
$E_{\chi\chi}=N_\varphi m_\varphi$.  In general, at the leading order
of $A_\varphi$, the decay rate of the coherent state is also obtained
by calculating the tree-level production rate of the final-state
particles treating the scalar condensation as an external field.  If
we consider higher order contributions, however, such a calculation
breaks down; treating the scalar condensation as the external field,
denominators of some propagators vanish in certain types of diagrams.
Notice that, in Eq.\ (\ref{Im(T)}), such a difficulty does not exist.
(See also the following discussion.)

\subsection{Small velocity limit}

In the previous subsection, we have calculated leading-order
contributions to the decay rates of each mode in the small amplitude
limit.  Calculations of the contributions which are higher order in
the amplitude are straightforward.  In this subsection, we discuss
when the small-amplitude expansion breaks down, taking the
$N_\varphi=1$ mode as an example.

If we calculate $O(|A_\varphi|^{2p})$ contributions to the decay rate
of such mode, which are from diagrams with $2p$ external $\varphi$
insertions, one finds that the imaginary part of the $T$-matrix element
is inversely proportional to the powers of $\beta_1$ in the
$m_\varphi\rightarrow 2m_\chi$ limit.  First, let us derive such a
behavior with explicit calculation.

\begin{figure}[t]
 \centerline{\epsfxsize=0.4\textwidth\epsfbox{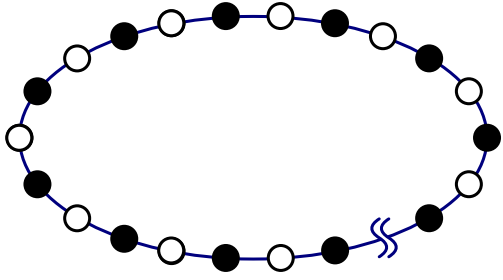}}
 \caption{\small Feynman diagrams which give the most singular
   behavior for the $N_\varphi=1$ process in the small $\beta_1$
   limit. The white dot $\circ$ indicates $\varphi_-$ insertion while
   the black dot $\bullet$ is for $\varphi_+$ insertion.}
 \label{fig:loop_smallb}
\end{figure}

For the $N_\varphi=1$ mode, the most important Feynman diagrams in the
$m_\varphi\rightarrow 2m_\chi$ limit are those in which $\varphi_+$
and $\varphi_-$ insertions are next to each other.  (See Fig.\
\ref{fig:loop_smallb}.)  As we will discuss, other types of diagrams
with fixed $p$ have less singular behavior when $\beta_1\rightarrow
0$.  We also note here that the diagram in Fig.\ \ref{fig:loop_smallb}
contributes only to the $N_\varphi=1$ mode.  (The imaginary part of
the $T$-matrix vanishes when $N_\varphi=0$.)

Taking into account the diagram shown in Fig.\ \ref{fig:loop_smallb},
the $T$-matrix element becomes
\begin{eqnarray}
 i {\cal T}^{(N_\varphi=1)}_{\beta_1\rightarrow 0} = L^3 T
  \frac{1}{2p} \left| \frac{\mu A_\varphi}{2} \right|^{2p}
  \int \frac{d^4 \tilde{k}}{(2\pi)^4}
  \left( \tilde{k}^2 - m^2_\chi + i 0^+ \right)^{-p}
  \left[ (\tilde{k} - Q_\varphi )^2 - m^2_\chi + i 0^+ \right]^{-p},
  \label{T(beta)}
\end{eqnarray}
and the imaginary part of ${\cal T}^{(N_\varphi=1)}_{\beta_1\rightarrow
0}$ is given by
\begin{eqnarray}
  \Im \left[ {\cal T}^{(N_\varphi=1)}_{\beta_1\rightarrow 0} \right] =
   - \frac{L^3 T}{32\pi} m_\varphi^4 \beta_1
  \left| \frac{\mu A_\varphi}{2} \right|^{2}
  \times
  \frac{(4p-7)!!}{p!(p-1)!} 
  \left| \frac{\mu A_\varphi}{2m_\varphi^2 \beta_1^2} \right|^{2(p-1)}
  + O(\beta_1^{7-4p}),
  \label{Im[T(beta)]}
\end{eqnarray}
where $(2p-1)!!\equiv\prod_{I=1}^{p}(2I-1)$ for $p\geq 1$, and
$(-3)!!\equiv -1$.  (For details, see Appendix \ref{app:smallb}.)
Hereafter, we neglect $O(\beta_1^{7-4p})$ contribution in Eq.\
(\ref{Im[T(beta)]}), and the decay rate of the coherent state
becomes
\begin{eqnarray}
  \Gamma^{(N_\varphi=1)}_{\beta_1\rightarrow 0} 
  =
  -\frac{1}{16\pi}
  m_\varphi^4 \beta_1
  \left| \frac{\mu A_\varphi}{2} \right|^{2}
  \sum_{p=1}^\infty
  \frac{(4p-7)!!}{p!(p-1)!} 
  \left| \frac{\mu A_\varphi}{2m_\varphi^2 \beta_1^2} \right|^{2(p-1)}.
  \label{Gamma_1(b->0)}
\end{eqnarray}

As discussed in Appendix \ref{app:smallb}, the inverse powers of
$\beta_1$ stems from the derivative of the function $B(m_\phi^2;
\xi_i, \xi_j)$ given in Eq.\ (\ref{fnB}) with respect to $\xi_i$ or
$\xi_j$.  The order of the derivatives is equal to the number of the
propagators whose denominators vanish in the on-shell limit
$\xi_i\rightarrow m_\chi^2$ ($i=1-2p$).  Number of such propagators is
maximized for the diagram given in Fig.\ \ref{fig:loop_smallb}.  Thus,
we safely neglect other types of diagrams in studying the case of
$\beta_1\rightarrow 0$.

As we mentioned, $\Gamma^{(N_\varphi=1)}_{\beta_1\rightarrow 0}$
becomes singular when $\beta_1\rightarrow 0$.  In other words, for the
$N_\varphi=1$ mode, the small-amplitude expansion breaks down when
$|A_\varphi|$ is comparable to $\mu^{-1}m_\varphi^2\beta_1^2$.  These
behaviors are related to the fact that instability bands appear in the
solution to the classical wave equation of the scalar field (i.e.,
$\chi$ in our argument) which couples to a oscillating scalar field
(i.e., $\varphi$).  With the interaction given in Eq.\ (\ref{A-term}),
the wave equation of $\chi$ (with the 3-momentum ${\bf k}$) in the
oscillating background is given by the Mathieu equation:
\begin{eqnarray}
  \frac{d^2 \chi_{\bf k}}{dt^2}
  + \left(
    {\bf k}^2 + m_\chi^2 + \mu |A_\varphi| \cos m_\varphi t
  \right)
  \chi_{\bf k} = 0.
\end{eqnarray}
Parametrizing the momentum of $\chi$ as
\begin{eqnarray}
  {\bf k}^2 = \frac{1}{4} m_\varphi^2
  \left( \beta_1^2 + \epsilon \right),
\end{eqnarray}
the lowest instability band in the small-amplitude limit is given by
\cite{Yoshimura:1995gc}
\begin{eqnarray}
  -\theta < \epsilon < \theta,
  \label{eps_pm}
\end{eqnarray}
where
\begin{eqnarray}
 \theta \equiv \frac{2 \mu |A_\varphi|}{m_\varphi^2}.
\end{eqnarray}
From the study of the parametric resonance, the momentum of $\chi$
produced by the decay of the scalar condensation is in the range
given by Eq.\ (\ref{eps_pm}).  For the consistency of the calculation,
the mass of the initial-state particle $\varphi$ should be large
enough so that ${\bf k}^2$ is positive even for $\epsilon\sim
-\theta$; otherwise, $O(\theta^2)$ contributions may be also
important.  This argument gives the limitation of the small-amplitude
approximation; $\beta_1\gg \theta^{1/2}$ is required, which
results in $|A_\varphi|\gg\mu^{-1}m_\varphi^2\beta_1^2$.

The above argument is supported by the fact that
$\Gamma^{(N_\varphi=1)}_{\beta_1\rightarrow 0}$ given in Eq.\
(\ref{Gamma_1(b->0)}) is equal to the decay rate obtained by the
parametric-resonance analysis.  Indeed,
$\Gamma^{(N_\varphi=1)}_{\beta_1\rightarrow 0}$ is also expressed as
\begin{eqnarray}
 \Gamma^{(N_\varphi=1)}_{\beta_1\rightarrow 0} &=&
  \frac{1}{2} m_\varphi
  \int_{|{\bf k}|_- <|{\bf k}| <|{\bf k}|_+}
  \frac{d^3 {\bf k}}{(2\pi)^3}
  \lambda^{(N_\varphi=1)} ({\bf k})
  \nonumber \\ &=&
  \frac{1}{128 \pi^2}
  m_\varphi^4
  \int_{-\theta}^{\theta} d \epsilon
  \sqrt{ \left(\beta_1^2 + \epsilon \right)
    \left(\theta^2 - \epsilon^2 \right)},
  \label{Gamma_1(Param)}
\end{eqnarray}
where, in the first equality, $|{\bf
  k}|_\pm\equiv\frac{1}{2}m_\varphi\sqrt{\beta_1^2\pm\theta}$, and
$\lambda^{(N_\varphi=1)}=\frac{1}{2}\sqrt{\theta^2-\epsilon^2}$ is the
``growth-rate factor'' for the $N_\varphi=1$ mode obtained in the
study of the Mathieu equation \cite{Yoshimura:1995gc}.  The
equivalence of Eq.\ (\ref{Gamma_1(b->0)}) and Eq.\
(\ref{Gamma_1(Param)}) can be seen by expanding $\sqrt{\beta_1^2 +
  \epsilon}$ in the integrand of Eq.\ (\ref{Gamma_1(Param)}) around
$\epsilon=0$ (assuming $\beta_1>\theta^{1/2}$).  Eq.\
(\ref{Gamma_1(Param)}) is nothing but the decay rate of the scalar
condensation in the small-amplitude limit derived from the
parametric-resonance analysis.

Eq.\ (\ref{Gamma_1(Param)}) (and Eq.\ (\ref{Gamma^N})) also shows the
fact that, at least at the small-amplitude limit, the results from the
parametric-resonance analysis is obtained in our procedure where the
quantum state describing the scalar condensation is postulated to be
the coherent state.  The equivalence of two approaches is also
expected from the fact that the basic equations governing the behavior
of the parametric resonance is derived in our framework.  In
particular, we can calculate the density matrix of the final-state
particle $\chi$ in the quantum field theory.  We can see that the
Mathieu equation shows up in the calculation and that the resultant
density matrix is the same as the one obtained in the study of
parametric resonance.  These subjects will be discussed in the next
subsection.

\subsection{Calculation of the density matrix}

In the classical treatment of the parametric resonant system, it is
well known that the Mathieu equation appears as the equation of motion
for the $\chi$ field. Thus, the equation is also expected to be
obtained in the quantum-field-theory treatment.  In this subsection,
we show the derivations of the Mathieu equation and the density matrix
of $\chi$ explicitly.

Since we are interested in the case where the scalar field $\varphi$
initially forms the scalar condensation oscillating around the minimum
of its potential, we describe the initial state (which is taken at
$t=0$ in this subsection) as
\begin{eqnarray}
  | i \rangle = | \varphi \rangle \otimes | 0 \rangle_{\chi},
\end{eqnarray}
where the first and second kets represent the states for $\varphi$ and
$\chi$, respectively.  In addition, $| 0 \rangle_{\chi}$ is the
vacuum of the $\chi$ field.  (In the following, we omit the subscript
$\chi$.)

For our argument, it is convenient to use the density matrix of the
total system in the Schroedinger picture. The density matrix at the
time $T$ is simply given by
\begin{eqnarray}
  \hat{\rho}_{\rm tot} (T)
  =
  e^{-i \hat{H} T}
  | i \rangle \langle i | e^{i \hat{H} T},
\end{eqnarray}
with $\hat{H}$ being the Hamiltonian of the total system.  We consider
the properties of $\hat{\rho}_{\rm tot}$ in the coordinate basis:
\begin{eqnarray}
  | q \otimes X \rangle \equiv
  | q \rangle  \otimes | X \rangle,
\end{eqnarray}
where $| q\rangle$ and $| X\rangle$ are eigenstates of the field
operators $\hat{\varphi}$ and $\hat{\chi}$, respectively:
\begin{eqnarray}
  \hat{\varphi}(t, {\bf x}) | q \rangle =
  q({\bf x})  | q \rangle,~~~
  \hat{\chi}(t, {\bf x}) | X \rangle =
  X({\bf x})  | X \rangle.
\end{eqnarray}
Then, the density matrix in the coordinate basis, $\rho_{\rm
  tot}[ q,X;q',X']\equiv\langle q\otimes X|\hat{\rho}_{\rm
  tot}(T)|q'\otimes X'\rangle$, is given by
\begin{eqnarray}
  \rho_{\rm tot} [ q,X ; q',X' ]
  &=&
  \int {\cal D} q_i \int {\cal D} q_i' 
  \int {\cal D} X_i \int {\cal D} X'_i
  \langle q_i | \varphi \rangle
  \langle X_i | 0 \rangle
  \langle \varphi | q'_i \rangle
  \langle 0 | X'_i \rangle
  \nonumber \\ &&
  K [ q,X ; q_i,X_i ] K^* [ q',X' ; q_i',X_i' ].
\end{eqnarray}
The kernel is represented in the path integral form as
\begin{eqnarray}
  K [ q,X ; q_i,X_i ]
  =
  \int^{\chi(T,{\bf x}) = X({\bf x})}_{\chi(0,{\bf x}) = X_i({\bf x})}
  {\cal D}\chi
  \int^{\varphi(T,{\bf x}) = q({\bf x})}_{\varphi(0,{\bf x}) = q_i({\bf x})}
  {\cal D}\varphi
  e^{i S_{\rm tot}},
  \label{kernel}
\end{eqnarray}
where $S_{\rm tot}$ is the total action.  For the explicit form of the
kernel, see \cite{Feynman:1963fq,FeynmanHibbs}.

In order to study the behavior of the $\chi$ field, we derive the
reduced density matrix of $\chi$ by tracing out $q$ and $q'$
variables:
\begin{eqnarray}
  \rho_{\rm red} [X ; X'] \equiv 
  \int {\cal D} q
  \rho_{\rm tot} [ q,X ; q,X' ].
\end{eqnarray}
For this purpose, we define
\begin{eqnarray}
  \tilde{q}_{\bf k} &\equiv&
  L^{-3/2} \int d^3 {\bf x} q({\bf x}) e^{-i{\bf kx}},
  \\
  \tilde{X}_{\bf k} &\equiv&
  L^{-3/2} \int d^3 {\bf x} X({\bf x}) e^{-i{\bf kx}}.
\end{eqnarray}
Then, with the use of the properties of the coherent state, we obtain
the following relations:
\begin{eqnarray}
  \langle q | \varphi \rangle &=& 
  \exp \left[ 
    -\frac{1}{2} m_\varphi (\tilde{q}_{\bf 0}-L^{3/2} A_\varphi)^2 \right]
  \prod_{{\bf k}\neq {\bf 0}}
  \exp \left[ -\frac{1}{2} E_{\bf k} |\tilde{q}_{\bf k}|^2 \right],
  \label{wf_varphi}
  \\
  \langle X | 0 \rangle &=&
  \prod_{\bf k}
  \exp \left[ -\frac{1}{2} \omega_{\bf k} |\tilde{X}_{\bf k}|^2 \right],
  \label{wf_chi}
\end{eqnarray}
where, in this subsection, we consider the case where $A_\varphi$ is
real.  ($A_\varphi$ can be taken to be real with a relevant shift of
the time variable.) In addition, $\omega_{\bf k}^2 = {\bf k}^2 +
m^2_\chi$.  In Eqs.\ (\ref{wf_varphi}) and (\ref{wf_chi}), we omit
unimportant numerical constants.  Using Eq.\ (\ref{wf_varphi}), the
reduced density matrix becomes
\begin{eqnarray}
  \rho_{\rm red} [X;X'] =
  \int {\cal D}X_i {\cal D} X'_i
  \langle X_i | 0 \rangle
  \langle 0 | X'_i \rangle
  \int^{\chi(T,{\bf x}) = X}_{\chi(0,{\bf x}) = X_i}
  {\cal D}\chi
  \int^{\chi'(T,{\bf x}) = X'}_{\chi'(0,{\bf x}) = X'_i}
  {\cal D}\chi'
  e^{i \bar{S} [\chi] - i \bar{S}[\chi'] + {\cal C}[\chi, \chi']},
\end{eqnarray}
where
\begin{eqnarray}
  \bar{S} [\chi] =
  \int_0^T dt \int d^3 {\bf x}
  \left[
    \frac{1}{2} \partial_\mu \chi \partial^\mu \chi
    - \frac{1}{2} m_\chi^2 \chi^2
    - \frac{1}{2} \mu A_\varphi \chi^2 \cos m_\varphi t 
  \right],
\end{eqnarray}
while ${\cal C}$ gives the collision terms:
\begin{eqnarray}
  {\cal C}[\chi, \chi']
  &=&
  - \frac{\mu^2}{4}
  \int_0^T dt \int_0^t dt' \int d^3 {\bf x} \int d^3 {\bf x}'
  \int \frac{d^3 {\bf k}}{ (2\pi)^3 2 E_{\bf k}}
  \cos {\bf k} \left( {\bf x} - {\bf x}' \right) 
  \nonumber \\ &&
  \left[ \chi^2 (x) - \chi^{\prime 2}(x) \right]
  \left[
    e^{-i E_{\bf k} (t-t')} \chi^2 (x')
    -
    e^{ i E_{\bf k} (t-t')} \chi^{\prime 2} (x')
  \right].
\end{eqnarray}
It can be seen that the periodic perturbation term, $\chi^2 \cos
m_\varphi t$, appears in the reduced density matrix. Collision terms,
which are proportional to $\chi^4$, $\chi^{\prime 4}$, and $\chi^2
\chi^{\prime 2}$, also appear after the integration. Since no
approximation was made to derive the density matrix, the above formula
can be used at any amplitude of the $\varphi$ field, $A_\varphi$.
Furthermore, we can derive the kinetic equation by calculating the
correlation function of the $\chi$ field on the reduced density
matrix, which allows us to describe the non-linear dynamics of the
$\chi$ system, as pointed out in \cite{Matsumoto:1999us}.

When the amplitude $A_\varphi$ is small enough, we can treat the
collision terms as perturbations. Then, the reduced density matrix is,
at the leading order calculation, written by using the wave functional
of the $\chi$ field:
\begin{eqnarray}
  \rho_{\rm red} \left[ X; X' \right] =
  \Psi[T, X] \times \Psi^*[T; X'],
\end{eqnarray}
where 
\begin{eqnarray}
  \Psi[T, X] = 
  \int {\cal D} X_i \langle X_i | 0 \rangle
  \int^{\chi(T,{\bf x})=X}_{\chi(0,{\bf x})=X_i} {\cal D}\chi 
  e^{i \bar{S}[\chi]}.
\end{eqnarray}
The wave functional is written as a product of an infinite set of
wave functions of harmonic oscillators as
\begin{eqnarray}
 \Psi[T, X] =
 \prod_{\bf k}
 \int^\infty_{- \infty} 
 d \tilde{X}_{i,{\bf k}}
 \exp \left[
   -\frac{1}{2} \omega_{\bf k} 
   |\tilde{X}_{i,{\bf k}}|^2
 \right]
 \int^{
   \tilde{\chi}_{\bf k}(T)=\tilde{X}_{\bf k}}_{
   \tilde{\chi}_{\bf k}(0)=\tilde{X}_{i,{\bf k}}}
 {\cal D} \tilde{\chi}_{\bf k}
  e^{i \bar{S}_{\bf k} \left[ \tilde{\chi}_{\bf k} \right]},
\end{eqnarray}
where
\begin{eqnarray}
  \bar{S}_{\bf k} \left[ \tilde{\chi}_{\bf k} \right] = 
  \int^T_0 dt
  \left[
    \frac{1}{2} |\dot{\tilde{\chi}}_{\bf k}|^2
    -
    \frac{1}{2}
    \left(
    {\bf k}^2 + m^2_\chi + \mu A_\varphi \cos m_\varphi t
    \right)
    |\tilde{\chi}_{\bf k}|^2
    \right],
  \label{S_k}
\end{eqnarray}
with the ``dot'' being the derivative with respect to time.
$\Psi[T,X]$ satisfies the boundary condition $\Psi[0,X]=\prod_{\bf
  k}e^{-\omega_{\bf k}|\tilde{X}_{{\bf k}}|^2/2}$ (up to
normalization), and its evolution is governed by the wave equation
derived from the action given in Eq.\ (\ref{S_k}).  Thus, $\Psi[T,X]$
is nothing but the wave functional obtained in
\cite{Yoshimura:1995gc}:
\begin{eqnarray}
  \Psi[T, X]
  =
  \prod_{\bf k}
  \frac{1}{\sqrt{u_{\bf k} (T)}}
  \exp
  \left[
    \frac{i}{2}
    \frac{\dot{u}_{\bf k} (T)}{u_{\bf k} (T)}
    |\tilde{X}_{\bf k}|^2
  \right],
\end{eqnarray}
where $u_{\bf k}(t)$ is the solution to the Mathieu equation:
\begin{eqnarray}
 \frac{d^2 u_{\bf k}}{dt^2}
 +
 \left( \omega_{\bf k}^2 + \mu A_\varphi \cos m_\varphi t
 \right) u_{\bf k}
 =
 0,
\end{eqnarray}
with the conditions $u_{\bf k}(0) = \sqrt{\pi/\omega_{\bf k}}$ and
$\dot{u}_{\bf k}(0) = i \sqrt{\omega_{\bf k} \pi}$.  This fact
supports that the scalar condensation is well described by the
coherent state in the quantum field theory.

Before closing this section, we emphasize that the quantity $\Im
[\langle\varphi|{\cal T}|\varphi\rangle]$ is relatively easily
calculated with wide variety of interactions and final-states.  Thus,
for some applications, our procedure is more powerful than the
approach using the Mathieu equation.

\section{Decay via Anomaly}
\label{sec:anomaly}
\setcounter{equation}{0}

Next, let us consider the case where a complex scalar field $\phi$ may
decay via chiral and conformal anomalies.  As the fundamental theory,
we expect that there exists chiral fermions which have gauge quantum
numbers and that the complex scalar field couples to the chiral
fermions through a Yukawa interaction.  To make our discussion
definite, we consider $SU(N_{\rm c})$ gauge interaction; chiral
fermions $Q_L$ and $Q_R^c$ are in fundamental and anti-fundamental
representations of $SU(N_{\rm c})$, respectively, while $\phi$ is
singlet.

The complex scalar field couples to chiral fermions $Q_L$ and $Q_R^c$
via the Yukawa interaction
\begin{eqnarray}
  {\cal L}_{\rm Yukawa} = -y (\phi Q_L Q_R^c + {\rm h.c.}).
\end{eqnarray}
It should be noted that, in this model, there exists anomalous $U(1)$
symmetry, which we call $U(1)_A$; charges of $\phi$, $Q_L$ and $Q_R^c$
are $1$, $-\frac{1}{2}$, and $-\frac{1}{2}$, respectively.

When the amplitude of $\phi$ is large, fermions $Q_L$ and $Q_R^c$
acquire Dirac mass.  Thus, when $m_\phi\ll y|\phi|$, effective mass of
the fermions are much larger than the mass of $\phi$.  In this case,
the decay process $\phi\rightarrow Q_LQ_R^c$ is expected to be
kinematically forbidden.

When $m_\phi\ll y|\phi|$, it is rather convenient to consider the
low-energy effective field theory by integrating out the fermions.
The relevant (light) fields in the low-energy effective field theory
are $\phi$ and gauge fields as far as the effective mass of the
fermions are much larger than $m_\phi$.  In the following, we
concentrate on such a case; thus, we assume that the inequality
$m_\phi\ll y|\phi(x)|$ always holds at any point of the trajectory of
$\phi$.

We first consider the effects of the operator induced by the 
chiral anomaly:
\begin{eqnarray}
  {\cal L}_{\rm eff} = - i \lambda_{\rm I} 
  \left( \ln \phi - \ln \phi^\dagger \right)
  F^{\mu\nu} \tilde{F}_{\mu\nu},
  \label{phiFF-dual}
\end{eqnarray}
where $F_{\mu\nu}$ is the field-strength tensor and
\begin{eqnarray}
  \tilde{F}^{\mu\nu} \equiv 
  \frac{1}{2} \epsilon^{\mu\nu\rho\sigma} F_{\rho\sigma}.
\end{eqnarray}
In Eq.\ (\ref{phiFF-dual}) and hereafter, summation over the adjoint
gauge index is implicit.  In addition,
\begin{eqnarray}
  \lambda_{\rm I} \equiv \frac{g^2}{32\pi^2} T_R,
\end{eqnarray}
with $g$ being the gauge coupling constant of $SU(N_{\rm c})$, and
$T_R=\frac{1}{2}$.

Now, let us discuss the decay of the coherent state given in Eq.\
(\ref{|complex>}).  At the leading order in $\lambda_{\rm I}$, which
is of $O(\lambda_{\rm I}^2)$ in the calculation of $\langle\phi
|\hat{\cal T}|\phi\rangle$, we obtain
\begin{eqnarray}
  \langle \phi | i \hat{\cal T} | \phi \rangle 
  &=&
  - \frac{1}{2} \lambda_{\rm I}^2 \int d^4 x d^4 x'
  \langle 0 | T \hat{F}_{\mu\nu} (x) \hat{\tilde{F}}^{\mu\nu} (x)
  \hat{F}_{\mu'\nu'} (x') \hat{\tilde{F}}^{\mu'\nu'} (x') | 0 \rangle
  \nonumber \\ &&
  \left[ \ln \phi (x) - \ln \phi^\dagger (x) \right]
  \left[ \ln \phi (x') - \ln \phi^\dagger (x') \right].
  \label{<phi|iT|phi>}
\end{eqnarray}

In the following, we consider the case that $A_\phi\geq A_{\bar{\phi}}$.
Then, we expand $\ln \phi (x)$ as
\begin{eqnarray}
 \ln \phi (x) = 
  \ln (A_\phi e^{-iQ_\phi x}) +
  \sum_{n=1}^\infty \frac{(-1)^{n-1}}{n}
  \left( \frac{A_{\bar{\phi}}^*}{A_\phi} \right)^{n}
  e^{2niQ_\phi x},
\end{eqnarray}
where $Q_\phi=(m_\phi,{\bf 0})$.  At the lowest order in $\lambda_{\rm
I}$, the decay rate of the coherent state can be obtained by calculating
the two-point functions of several types of operators with relevant
momentum injection.  For the local operator $\hat{\cal O}(x)$, let us
define
\begin{eqnarray}
 {\cal I}_{\cal O} (Q) \equiv
  -i \int d^4 x_1 d^4 x_2
  \langle 0 | T \hat{\cal O} (x_1) \hat{\cal O} (x_2) | 0 \rangle 
  e^{iQ(x_1-x_2)}.
\end{eqnarray}
Then, by using the fact that $F_{\mu\nu}\tilde{F}^{\mu\nu}$ is
expressed as a total derivative:
\begin{eqnarray}
  F_{\mu\nu}\tilde{F}^{\mu\nu} = \partial_\mu K^\mu
  = \frac{1}{2} \partial_\mu 
  \left[ \epsilon^{\mu\nu\rho\sigma} A_\nu (\partial_\rho A_\sigma)
    + (\mbox{gauge field})^3 \right],
\end{eqnarray}
Eq.\ (\ref{<phi|iT|phi>}) becomes
\begin{eqnarray}
 \langle \phi | \hat{\cal T} | \phi \rangle =
  - 2 \lambda_{\rm I}^2 {\cal I}_{Q_\phi^\mu K_\mu} (0)
  - \lambda_{\rm I}^2 \sum_{n=1}^{\infty}
  \frac{1}{n^2} \left| \frac{A_{\bar{\phi}}}{A_\phi} \right|^{2n}
  {\cal I}_{F_{\mu\nu}\tilde{F}^{\mu\nu}} (2nQ_\phi).
\end{eqnarray}

It is notable that ${\cal I}_{Q_\phi^\mu K_\mu}(0)$ has no imaginary
part because there is no momentum injection into the internal
gauge-boson lines from the $Q_\phi^\mu K_\mu$-vertex.  Thus, the
coherent state does not decay if $A_\phi=0$ or $A_{\bar{\phi}}=0$.
(This statement holds even after taking into account the higher order
terms in $\lambda_{\rm I}$.)  This fact can be understood by the
conservation of the $U(1)_A$ charge.  With a fixed value of the total
energy of the system, $U(1)_A$ charge is maximized when $A_\phi=0$ or
$A_{\bar{\phi}}=0$.  Thus, if $U(1)_A$ charge is conserved, the decay of
$\phi$ in the condensation into the gauge bosons is
forbidden.  Of course, the interaction given in Eq.\
(\ref{phiFF-dual}) breaks $U(1)_A$ symmetry because ${\cal L}_{\rm eff}$
is not invariant under the $U(1)_A$ transformation. This is
due to the fact that ${\cal L}_{\rm eff}$ is induced by the chiral
anomaly.  However,  we can add new fermions, which
we call $Q'_L$ and $Q'^c_R$, to have conserved $U(1)$ symmetry.
Indeed, with $Q'_L$ and $Q'^c_R$, which are in fundamental and
anti-fundamental representation of $SU(N_{\rm c})$, respectively, we
can define non-anomalous $U(1)_A$ symmetry by assigning charge
$+\frac{1}{2}$ to both of $Q'_L$ and $Q'^c_R$.  (Notice that $Q'_L$
and $Q'^c_R$ do not have to couple to $\phi$.)  In this case,
conservation of the $U(1)_A$ charge is obvious and the decay of $\phi$
into the gauge bosons is completely forbidden.  Thus, the $U(1)_A$ charge
stored in the scalar condensation cannot be released by the
interaction given in Eq.\ (\ref{phiFF-dual}).  This fact may have some
relevance in the study of the decay of scalar condensations in various
cosmological scenarios, in particular, in the Affleck-Dine scenario
\cite{Affleck:1984fy}.  In the absence of $Q'_L$ and $Q'^c_R$,
instanton effects may generate new interactions which explicitly
breaks $U(1)_A$ symmetry.  In such a case, decay of the coherent state
occurs via such new interactions.

Using the relation
\begin{eqnarray}
 \Im \left[ {\cal I}_{F_{\mu\nu}\tilde{F}^{\mu\nu}} (Q) \right]
  = - \frac{N_{\rm c}^2 - 1}{4\pi} \left( Q^2 \right)^2 L^3 T,
\end{eqnarray}
the decay rate is given by
\begin{eqnarray}
 \Gamma_{F\tilde{F}} =
  \frac{{\rm Prob} (|\phi\rangle \rightarrow {\rm all})}{L^3 T} =
  \frac{8}{\pi} (N_{\rm c}^2 - 1) \lambda_{\rm I}^2 m_\phi^4 
  \sum_{n=1}^{\infty} n^2
  \left| \frac{A_{\bar{\phi}}}{A_\phi} \right|^{2n}.
  \label{Gamma_FFdual}
\end{eqnarray}
Since the decay rate vanishes if $A_\phi=0$ or $A_{\bar{\phi}}=0$, the
decay of the coherent state in this case should be understood as an
annihilation between $\phi$ and $\bar{\phi}$ in the condensation;
same number of $\phi$ and $\bar{\phi}$ annihilate into the gauge boson
pair.

In the study of the decay of coherent state, the energy-loss rate is
also important.  Using the fact that the imaginary part of ${\cal
  I}_{F_{\mu\nu}\tilde{F}^{\mu\nu}}(2nQ_\phi)$ is from the decay
process into two gauge bosons with the total energy of $2nm_\phi$, the
energy-loss rate can be calculated.  So far, we have considered the
case where $A_\phi\geq A_{\bar{\phi}}$.  However, the decay rate for
the case of $A_\phi\leq A_{\bar{\phi}}$ is derived by interchanging
$A_\phi\leftrightarrow A_{\bar{\phi}}$ in the result.  Thus, we obtain
\begin{eqnarray}
 \left[ \frac{d \rho_\phi}{dt} \right]_{F\tilde{F}} =
  -\frac{16}{\pi} (N_{\rm c}^2 - 1) \lambda_{\rm I}^2 m_\phi^5
  \sum_{n=1}^{\infty} n^3
  \left[ \frac{\min(n_\phi, n_{\bar{\phi}})}
   {\max(n_\phi, n_{\bar{\phi}})}  \right]^{n},
  \label{drho_FFdual}
\end{eqnarray}
where we have used the fact that the number densities of $\phi$ and
$\bar{\phi}$ are proportional to $|A_\phi|^2$ and $|A_{\bar{\phi}}|^2$,
respectively.  (See Eqs.\ (\ref{n_phi}) and (\ref{n_phibar}).)  One may
simplify the above energy-loss rate by using
\begin{eqnarray}
 \sum_{n=1}^{\infty} n^3 r^n = \frac{r(1 + 4r + r^2)}{(1-r)^4}.
\end{eqnarray}

Here, we emphasize that the results given in Eqs.\
(\ref{Gamma_FFdual}) and (\ref{drho_FFdual}) can be used for any value
of the amplitude (as far as the effective mass of $Q_L$ and $Q_R^c$
are much larger than $m_\phi$).  When $n_\phi\gg n_{\bar{\phi}}$ or
$n_\phi\ll n_{\bar{\phi}}$, which corresponds to the case where the
classical motion of the scalar condensation is almost circular, the
energy-loss rate is well approximated by the leading term in Eq.\
(\ref{drho_FFdual}).  On the contrary, in the limit of
$n_{\bar{\phi}}\rightarrow n_\phi$, higher order terms become
important and the energy-loss rate is enhanced.  In this case,
however, one should note that, at some point of the classical
trajectory, $|\phi|$ approaches to the origin.  Then, the effective
mass of the fermions $Q_L$ and $Q_R^c$ may become so small that the
effective field theory, which is obtained by integrating out these
fermions, may break down.  We also note that, with the ratio
$n_\phi/n_{\bar{\phi}}$ being fixed, the decay and energy-loss rates
are independent of the amplitude of the scalar condensation.

Before closing this section, we also present the result for the case
where the scalar field $\phi$ couples to the gauge field as
\begin{eqnarray}
 {\cal L}_{\rm eff} = \lambda_{\rm R} 
  \left( \ln \phi + \ln \phi^\dagger \right)
  F^{\mu\nu} F_{\mu\nu}.
  \label{phiFF}
\end{eqnarray}
This type of interaction is also generated by integrating out particles
which acquire masses from the condensation of $\phi$ (like $Q_L$ and
$Q_R^c$).  At the leading order in $\lambda_{\rm R}$, the energy-loss
rate is given by
\begin{eqnarray}
 \left[ \frac{d \rho_\phi}{dt} \right]_{FF} =
  -\frac{16}{\pi} (N_{\rm c}^2 - 1) \lambda_{\rm R}^2 m_\phi^5
  \sum_{n=1}^{\infty} n^3
  \left[ \frac{\min(n_\phi, n_{\bar{\phi}})}
   {\max(n_\phi, n_{\bar{\phi}})}  \right]^{n}.
  \label{drho_FF}
\end{eqnarray}

\section{Summary}
\label{sec:summary}
\setcounter{equation}{0}

In this paper, we have discussed the decay processes of the scalar
condensation.  We postulated that the quantum state corresponding to
the scalar oscillation is the so-called coherent state in the quantum
field theory.  Then, by using the $S$-matrix unitarity, we have
developed the method to calculate the decay rate of the coherent
state.  We believe that our procedure can be applied to a large class
of models which may contain various types of interactions.

Then, in order to demonstrate how the decay rate is calculated, we
considered two examples.  First, we studied the case where the scalar
field $\varphi$ couples to another scalar field $\chi$ via three-point
interaction.  Using the small-amplitude approximation, we have
calculated the decay rate for the process where $N_\varphi$
($N_\varphi=1,2,3,\cdots$) of $\varphi$ in the condensation
simultaneously annihilate into a pair of $\chi$.  For the case of
$N_\varphi=1$, we have seen that the result is the same as that in the
conventional approach where the decay rate of the scalar condensation
is estimated by the product of the decay rate of single $\varphi$ in
the vacuum and the number density of $\varphi$.  We have also pointed
out that the small-amplitude approximation breaks down when the
amplitude becomes close to $\mu^{-1}m_\varphi^2\beta_1^2$, where $\mu$
is the coupling constant and $\beta_1$ is the velocity of $\chi$ in
the $N_\varphi=1$ mode.  Such a behavior is also expected from the
discussion based on the parametric-resonance.  Indeed, our procedure
reproduced the decay rate of the scalar condensation calculated from
the parametric-resonance analysis.

The second example was the case where the complex scalar field decays
into gauge bosons via the interaction induced by the chiral anomaly.
We have considered the case where the scalar potential has $U(1)_A$
symmetry to rotate $\phi\rightarrow e^{i\alpha}\phi$ at the classical
level, which is broken by the effect of the chiral anomaly.  In this
case, we could calculate the decay rate without using the
small-amplitude approximation.  We have seen that the decay process is
forbidden unless both the particle $\phi$ and its anti-particle
$\bar{\phi}$ exist in the condensation, and that the ``decay'' of the
coherent state is due to the annihilation between them.  Thus,
$U(1)_A$ charge stored in the condensation cannot be released by the
effective interaction induced by the chiral anomaly.

In our analysis, the effects of the cosmic expansion were completely
neglected.  However, we believe that our results are applicable to the
cosmological discussion as far as the expansion rate of the universe
is smaller than the mass of the scalar condensation.

{\it Acknowledgement}: This work was supported in part by the
Grant-in-Aid for Scientific Research from the Ministry of Education,
Science, Sports, and Culture of Japan, No.\ 19540255 (T.M.).

\appendix

\section{Calculation of the Imaginary Part}
\label{app:im}
\setcounter{equation}{0}

Although the technique which will be explained here is well-known, in
this appendix, we show how Eq.\ (\ref{Im(T)}) is derived for the sake
of some of the readers.  For this purpose, we calculate the imaginary
part of the following quantity:
\begin{eqnarray}
  I_{\cal F} (Q_\varphi) \equiv
  - i \int \frac{d^4 \tilde{k}}{(2\pi)^4}
  \prod_{I=1}^{2p}
  ( k_I^2 - m^2_\chi + i 0^+ )^2.
\end{eqnarray}
Here,
\begin{eqnarray}
  k_I \equiv \tilde{k} + \sum_{J=1}^{I} \varepsilon_{J} Q_\varphi,
\end{eqnarray}
where $\varepsilon_I=\pm 1$ (with $\varepsilon_1 + \cdots +
\varepsilon_{2p}=0$), and $Q_\varphi=(m_\varphi, {\bf 0})$.

In order to calculate the imaginary part of $I_{\cal F}$, it is 
convenient to rewrite $I_{\cal F}$ as
\begin{eqnarray}
  I_{\cal F} =
  - i \lim_{\xi\rightarrow m_\chi^2}
  \int \frac{d^4 \tilde{k}}{(2\pi)^4}
  \prod_{I=1}^{2p}
  ( k_I^2 - \xi_I + i 0^+ )^{-1},
  \label{I_F}
\end{eqnarray}
where the limit $\xi\rightarrow m_\chi^2$ indicates that
$\xi_I\rightarrow m_\chi^2$ ($I=1-2p$); before taking the limit,
$\xi_I$ are all set to be different.  Integrand of Eq.\ (\ref{I_F})
has poles at
$\tilde{k}_0=-m_\varphi\sum_{J=1}^I\varepsilon_J\pm\sqrt{{\bf
    k}_I^2+\xi_I}$ ($I=1-2p$) and, after $\tilde{k}_0$-integration,
$I_{\cal F}$ becomes
\begin{eqnarray}
  I_{\cal F} &=&
  - \pi \lim_{\xi\rightarrow m_\chi^2}
  \sum_{i=1}^{2p}
  \int \frac{d^3 \tilde{\bf k}}{(2\pi)^4}
  \left[ \frac{1}{2\tilde{k}_0}
    \prod_{I\neq i} ( k_I^2 - \xi_I + i 0^+ )^{-1}
    \right]_{\tilde{k}_0=
    -m_\varphi\sum_{J=1}^I\varepsilon_J+\sqrt{{\bf k}_I^2+\xi_I}}
  \nonumber \\ &=&
  - \pi \lim_{\xi\rightarrow m_\chi^2}
  \sum_{i=1}^{2p}
  \int \frac{d^4 \tilde{k}}{(2\pi)^4}
  \delta (k_i^2 - \xi_i)
  \prod_{I\neq i} ( k_I^2 - \xi_I + i 0^+ )^{-1},
\end{eqnarray}
where, in the second equality, the $\tilde{k}_0$-integration is
performed in the region where $k_{i0}\geq 0$.  Using the relation
$(x+i0^+)^{-1}=P(x^{-1})-i\pi\delta (x)$ (where ``$P$'' is for the
principal value), we obtain
\begin{eqnarray}
 \Im \left[ I_{\cal F} \right] =
  2 \pi^2 \lim_{\xi\rightarrow m_\chi^2}
  \sum_{i=1}^{2p-1} \sum_{j=i+1}^{2p}
  \int \frac{d^4 \tilde{k}}{(2\pi)^4}
  \delta (k_i^2 - \xi_i) \delta (k_j^2 - \xi_j)
  \prod_{I\neq i,j} ( k_I^2 - \xi_I )^{-1}.
\end{eqnarray}
Substituting the above expression into Eq.\ (\ref{T(2p)}), we
obtain Eq.\ (\ref{Im(T)}).

With the quantity
$N_\varphi=\left|\sum_{I=i+1}^{j}\varepsilon_I\right|$, $k_i$ and
$k_j$ are related as $k_i=k_j+N_\varphi Q_\varphi$ (or
$k_i=k_j-N_\varphi Q_\varphi$).  Thus, if $N_\varphi=0$, $k_i=k_j$ and
the imaginary part vanishes.  Notice also that the
constraints from the $\delta$-functions can be solved; by shifting the
integration variable $\tilde{k}$, $k_i$ and $k_j$ can be taken to be
$\tilde{k}$ and $\tilde{k}-N_\varphi Q_\varphi$, respectively.  Then,
the constraints from the $\delta$-functions become
$\tilde{k}Q_\varphi=\frac{1}{2}N_\varphi m_\varphi^2$.  Consequently,
the product $\prod_{I\neq i,j} ( k_I^2 - \xi_I )^{-1}$ becomes
$\tilde{k}$-independent.  The remaining part is proportional to the
two-body phase space for the process where the parent particle with
mass $N_\varphi m_\varphi$ decays into two daughter particles with
masses $\xi_i^{1/2}$ and $\xi_j^{1/2}$:
\begin{eqnarray}
  \int \frac{d^4 \tilde{k}}{(2\pi)^4}
  \delta (k_i^2 - \xi_i) \delta (k_j^2 - \xi_j)
  = \frac{1}{32 \pi^3}
  B ( N_\varphi^2 m_\varphi^2; \xi_i, \xi_j ),
\end{eqnarray}
where, for $\sqrt{Q^2}>\xi_i^{1/2}+\xi_j^{1/2}$,
\begin{eqnarray}
  B ( Q^2; \xi_i, \xi_j ) \equiv
  \frac{1}{Q^2} \sqrt{ \left( Q^2 \right)^2 
  - 2 (\xi_i + \xi_j) Q^2 + (\xi_i - \xi_j)^2 },
  \label{fnB}
\end{eqnarray}
while $B ( Q^2; \xi_i, \xi_j )=0$ for
$\sqrt{Q^2}\leq\xi_i^{1/2}+\xi_j^{1/2}$.  Notice that the function $B$
is related to $\beta_{N_\varphi}$ given in Eq.\ (\ref{beta_N}) as
\begin{eqnarray}
  \beta_{N_\varphi}
  = B ( N_\varphi^2 m_\varphi^2; m_\chi^2, m_\chi^2).
\end{eqnarray}

\section{Derivation of Eq.\ (\ref{Im[T(beta)]})}
\label{app:smallb}
\setcounter{equation}{0}

In this Appendix, we calculate the imaginary part of the following
integral
\begin{eqnarray}
  I_{{\rm Fig.}\ref{fig:loop_smallb}}
   = -i \int \frac{d^4 \tilde{k}}{(2\pi)^4}
  \left( \tilde{k}^2 - m^2_\chi + i 0^+ \right)^{-p}
  \left[ (\tilde{k} - Q_\varphi )^2 - m^2_\chi + i 0^+ \right]^{-p},
\end{eqnarray}
to derive Eq.\ (\ref{Im[T(beta)]}) from Eq.\ (\ref{T(beta)}).
Using the procedure given in Appendix \ref{app:im}, we express 
$I_{{\rm Fig.}\ref{fig:loop_smallb}}$ as
\begin{eqnarray}
 I_{{\rm Fig.}\ref{fig:loop_smallb}} =
  -i \lim_{\xi^{(\prime)}\rightarrow m_\chi^2}
  \int \frac{d^4 \tilde{k}}{(2\pi)^4}
  \prod_{I=1}^p
  \left( \tilde{k}^2 - \xi_I + i 0^+ \right)^{-1}
  \prod_{J=1}^p
  \left[ (\tilde{k} - Q_\varphi )^2 - \xi'_J + i 0^+ \right]^{-1}.
\end{eqnarray}
The imaginary part of this quantity is obtained with
\begin{eqnarray}
 \prod_{I=1}^p
  \left( \tilde{k}^2 - \xi_I + i 0^+ \right)^{-1}
  \rightarrow
  -i \pi \sum_{i=1}^p
  \prod_{I\neq i}
  ( \xi_i - \xi_I )^{-1}
  \delta \left( \tilde{k}^2 - \xi_i \right),
\end{eqnarray}
and with the similar replacement of the second product.  Then, the
imaginary part of $I_{{\rm Fig.}\ref{fig:loop_smallb}}$ becomes
\begin{eqnarray}
  \Im \left[ I_{{\rm Fig.}\ref{fig:loop_smallb}} \right]
  =
  \frac{1}{16\pi}
  \lim_{\xi^{(\prime)}\rightarrow m_\chi^2}
  \sum_{i=1}^p
  \sum_{j=1}^p
  \prod_{I\neq i}
  ( \xi_i - \xi_I )^{-1}
  \prod_{J\neq j}
  ( \xi'_j - \xi'_J )^{-1}
  B(m_\varphi^2; \xi_i, \xi'_j).
\end{eqnarray}
We can use the relation:
\begin{eqnarray}
  \lim_{x_1\rightarrow x} \cdots
  \lim_{x_p\rightarrow x}
  \sum_{i=1}^p f(x_i)
  \prod_{j\neq i} (x_i-x_j)^{-1}
  = \frac{1}{(p-1)!}
  \frac{d^{p-1}}{dx^{p-1}} f(x),
\end{eqnarray}
to obtain
\begin{eqnarray}
  \Im \left[ I_{{\rm Fig.}\ref{fig:loop_smallb}} \right]
  = 
  \frac{1}{16\pi}
  \frac{1}{[(p-1)!]^2}
  \left[
    \frac{\partial^{(p-1)}}{\partial \xi^{(p-1)}}
    \frac{\partial^{(p-1)}}{\partial \xi'^{(p-1)}}
    B(m_\varphi^2; \xi, \xi')
    \right]_{\xi=\xi'=m_\chi^2}.
\end{eqnarray}
Taking the $O(\beta_1^{1-4(p-1)})$ term from the above expression,
which is the most singular one when $\beta_1\rightarrow 0$, 
Eq.\ (\ref{Im[T(beta)]}) is derived.

\end{document}